\documentclass[aps,prb,twocolumn,showpacs]{revtex4}
\usepackage{ifthen}
\newcommand{\usepdfs}{false}
\ifthenelse{\equal{\usepdfs}{true}}{
  \usepackage[pdftex]{graphicx}
}
{
  \usepackage[dvips]{graphicx}
}
\ifthenelse{\equal{\usepdfs}{true}}{
  
}
{
  
}

\usepackage{amssymb,amsmath,amsfonts,hyperref}
\usepackage{wasysym}
\usepackage{latexsym}
\usepackage{subfigure}

\renewcommand{\u}{\uparrow}

\newcommand{\grad}{\nabla}

\newcommand{\chromite}{ACr$_2$O$_4$}
\newcommand{\znaf}{ZnCr$_2$O$_4$}
\newcommand{\cdaf}{CdCr$_2$O$_4$}
\newcommand{\hgaf}{HgCr$_2$O$_4$}

\begin{document}

\title{Mechanisms of degeneracy breaking in pyrochlore antiferromagnets}
\date{\today} 

\author{Doron L. Bergman$^1$, Ryuichi Shindou$^1$, Gregory
  A. Fiete$^2$, and Leon Balents$^1$}
\affiliation{${}^1$Department of Physics, University of California,
  Santa Barbara, CA 
93106-9530\\$^2$Kavli Institute for Theoretical Physics, University of 
California, Santa Barbara, CA 93106-4030}

\begin{abstract}

  Motivated by the low temperature magnetization curves of several
  spinel chromites, we theoretically study classical mechanisms of
  degeneracy lifting in pyrochlore antiferromagnets.  Our main focus
  is on the coupling of spin exchange to lattice distortions.  Prior
  work by Penc et al. (Phys. Rev. Lett. 93, 197203 (2004)) has 
  demonstrated that such coupling leads to a
  robust magnetization plateau at half the saturation moment per spin,
  in agreement with experiment.  We show that a simple Einstein model
  incorporating local site distortions generates a ``universal''
  magnetic order on the plateau, and highlight the distinct
  predictions of this model from that in Penc et al. (Phys. Rev. Lett. 93, 197203 (2004)).  We
  also consider the complementary degeneracy-lifting effects of
  further neighbor exchange interactions.  We discuss the implications
  for transitions off the plateau at both the high field and low field
  end, as well as at fields close to the saturation value.  
  We predict
  that under certain circumstances there is spontaneous {\sl uniform}
  XY magnetization (transverse to the field) for field values just
  above the plateau.  These features may be tested in experiments.
  While selecting a unique magnetic order in the half magnetization plateau,
  at zero magnetic field the Einstein model retains
  an extensive degeneracy, though significantly reduced compared with the 
  pure Heisenberg antiferromagnet. 
  
\end{abstract}
\date{\today}
\pacs{75.10.-b,75.10.Jm,75.25.+z}



\maketitle

\section{Introduction}\label{intro}

The pyrochlore lattice with nearest-neighbor anti-ferromagnetically
coupled spins is well-known as one of the most frustrated and
degenerate magnetic systems.\cite{Moessner:cjp01,Moessner:prb98}
Ultimately, this degeneracy must be lifted at low temperature, but the
mechanisms responsible can vary greatly from material to material and
also depend on applied fields, pressure, and other variables.  In this
paper we focus primarily on how degeneracies present at finite magnetic
fields are lifted by the coupling of spin and lattice degrees of
freedom.

Recent experiments on a number of insulating chromite compounds, namely
\znaf, \cdaf\, and \hgaf\, have revealed distinctive common features in
their low--temperature magnetization curves\cite{Ueda:prb06,Ueda:prl05}
and other interesting properties in neutron
scattering.\cite{Lee:nat02,chung:prl05} At low magnetic fields the
magnetization curve grows linearly with magnetic field.  At one point
there is a sharp jump in magnetization onto a rather robust plateau,
with half the full magnetization per spin. In \hgaf\, it is possible at
yet higher fields to observe a smooth transition off of the
half--magnetization plateau, and a gradual increase in magnetization up
to what may be a fully polarized plateau state. \cite{Ueda:prb06}

The Cr$^{+3}$ ions sit at the center of octahedra of O$^{-2}$ ions, and 
thus the outer d--orbital electron shell undergoes crystal field
splitting to a lower energy $t_{2g}$ orbital triplet, and a higher
energy $e_g$ orbital doublet. The $t_{2g}$ orbitals hold 3 electrons,
and therefore by Hund's rule form a spin $\frac{3}{2}$ degree of
freedom, with no orbital degeneracy (therefore the cooperative
Jahn--Teller effect cannot lift the degeneracy in this system).  These spins
are the source of magnetic behavior in these compounds.  The Cr$^{+3}$
ions sit on the sites of a pyrochlore lattice, and therefore a minimal
model for the magnetic properties of these compounds is the
nearest--neighbor Heisenberg anti--ferromagnet, with the
Hamiltonian 
\begin{equation}
{\mathcal H} = J \sum_{\langle i j \rangle} {\bf S}_i \cdot {\bf S}_j -
{\bf H} \cdot \sum_j {\bf S}_j 
\; .
\end{equation}
Here ${\bf H}$ is proportional to the applied magnetic (Zeeman) field.
In this paper, we will treat the spins as classical, and simplify
notation by normalizing them as unit vectors (absorbing a factor of
$S^2$ into $J$).  Much previous work has been devoted to this and
similar models in zero magnetic 
field.\cite{Hizi:prb06,Henley:prl06,Hizi:prl05,Elhajal:prb05,Koga:prb01,Harris:jap91,Tsunetsugu:prb01}  
A useful rewriting of the Hamiltonian is
\begin{equation}\label{Heisenberg}
{\mathcal H} =  \frac{J}{2} \sum_t \left[ ({\bf S}_t - {\bf h})^2  - {\bf h}^2 \right]
\; ,
\end{equation}
where ${\bf S}_t = \sum_{j \in t} {\bf S}_j$ is the sum of spins on a
tetrahedron labeled by $t$, ${\bf h} = {\bf H}/2J = h {\hat z}$ and we
have ignored a trivial constant term in the Hamiltonian.  This model has a
macroscopically degenerate classical ground state manifold at all fields up to full
polarization: any state with ${\bf S}_t={\bf h}$ on all tetrahedra is a
classical ground state.

\begin{figure}
	\centering
    \includegraphics[width=2.5in]{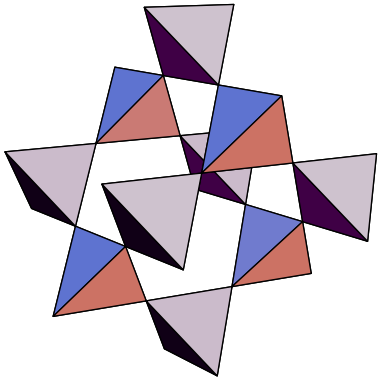}
	\caption{The pyrochlore lattice. A network of corner sharing tetrahedra.}
	\label{fig:Pyro_cell1}
\end{figure}

Within this (over-)simplified picture, the magnetization is everywhere a
smooth (linear at low temperatures, since the magnetization is
proportional to the average ${\bf S}_t$) function of the field, and
there are no plateaus.  Instead, some other effects or interactions must
be considered to explain the observed plateau.  On general grounds, a
plateau is expected to be associated with some breaking of degeneracy,
into a state in which the spontaneous static moments of the spins are
all parallel or antiparallel to the field axis.  Quantum mechanically,
this is simply because,
in a non-collinear {\sl ordered}
state, 
the Goldstone theorem associated with $U(1)$ symmetry-breaking around
the applied field direction ensures the presence of a gapless magnetic
excitation (spin wave), which contradicts the incompressibility of the
magnetization plateau.
The analogous classical argument is that a non-collinear spin state may
always be arbitrarily slightly deformed to lower its free energy in
response to a change in field, which by thermodynamics implies a
non-constant magnetization.  Hence, to understand the plateau, we seek
mechanisms to select one or a set of collinear ground states out of the
classically degenerate manifold.  On the pyrochlore lattice, the natural
collinear states for the half--polarized plateau
\cite{Ueda:prl05,Bergman:prl05,Bergman:prb05} are those with 3
``majority'' spins aligned parallel to the magnetic field, and 1
``minority'' spin aligned antiparallel to the field, {\sl on each
  tetrahedron}.  However, even if one assumes a collinear state for the
spins, 
a massive degeneracy of the ground state still remains, since there is
considerable freedom in fixing the location of the minority spin on each
tetrahedron.

The possibility that quantum fluctuations might control the state
selection -- of and within the collinear 3:1 manifold -- has been
explored elsewhere.\cite{Bergman:prl05,Bergman:prb05,Bergman:06} Here we will
investigate alternative mechanisms, within classical
models.\cite{Jia:prb05} A guide to the possible physical processes
involved comes from two sets of observations.  First, it has been noted
experimentally that the above chromite materials, \chromite, exhibit
strong magnetostriction, especially upon entering the plateau
region.\cite{Ueda:prl05,Ueda:prb06} This strongly suggests that
spin-lattice coupling plays an important role in the plateau formation.
Second, studies of the structurally and electronically analogous set of
spinels, ACr$_2$(S,Se)$_4$ -- with S or Se replacing O atoms and the
same non-magnetic A atoms -- display {\sl ferromagnetic} tendencies or
long-range order, and in some cases an apparent competition of
ferromagnetic and antiferromagnetic interactions.\cite{VanStepele:note}
This indicates drastic changes in the magnetic interactions may be
effected by small changes in structural parameters.  More specific
implications of the trends in these materials for the chromites will be
discussed later. 

This main analysis and results of this paper are as follows.  Guided
by the above observations, we focus primarily on a {\sl minimal} model
for the plateau structure involving {\sl only} spin-lattice coupling.
In this minimal model, the lattice modes are taken into account by the
simplest possible Einstein phonons describing motions of the magnetic
sites.  We show that this model indeed captures a simple and robust
mechanism for plateau formation {\sl and} predicts a unique ordered
3:1 state -- the ``R'' state, shown in Fig.\ref{fig:R_Config} -- on
the plateau.  Extended to the full range of magnetic fields, this
Einstein model predicts a first-order transition to a non-collinear
ground state at lower fields below the plateau, and a second-order
transition to a {\sl canted ferrimagnetic} state above the plateau.
The canted ferrimagnet retains the Ising order of the R state, but in
addition possesses XY \emph{ferrimagnetic} order of the magnetic
moments transverse to the field axis.  At zero field, the Einstein
model retains a large ground state degeneracy, though it is
still vastly reduced from that of the ideal model without spin-lattice
coupling.

A plateau with the same R state structure can also be stabilized by a
{\sl combination} of spin-lattice and further-neighbor exchange
interactions.  We give the conditions on these further-neighbor
exchanges for this to occur.  Consistency of this more complicated but
still feasible scenario could be then tested by placing independent
constraints on these couplings from other measurements.  This is
considered further in the Discussion.  

A number of studies of spin-lattice and further-neighbor exchange
effects in pyrochlores have already been carried out.  A well-known
analysis of certain zero field spin-lattice couplings by Tchernyshyov
{\sl et al.} christened the resulting degeneracy-breaking a ``spin Jahn
Teller'' effect.\cite{Tchernyshyov:prb02} Because this analysis was at
zero field, and because it considered only $q=0$ phonons, it has little
bearing on the present work.  More relevant is the pioneering study of
spin-lattice couplings on the plateau by Penc {\sl et al}.
\cite{Penc:prl04} Their work provides a simple explanation of the
plateau formation, but unlike the theory in this paper, does not explain
the breaking of degeneracy within the 3:1 states.  We will compare their
``bond phonon'' Hamiltonian with our Einstein model throughout this paper.

The remainder of this paper is organized as follows.  
In Sec.~\ref{sec:distortions} we present two models of 
spin-lattice coupling, a ``bond'' model and a ``site'' model.  
In Sec.~\ref{sec:away-from-half} we discuss the implications 
of these two models for the magnetic order on the half-saturation 
magnetization plateau and the transition off the high and low field 
edges of the plateau. 
We discuss a more general model with further neighbor spin 
interactions in Sec.~\ref{sec:J23}.  
Finally, a discussion 
of our main results and their relevance to experiment 
is given in Sec.~\ref{sec:discuss}. 

\section{Spin-lattice coupling}
\label{sec:distortions}

In this section we discuss some simple models for the coupling of the
magnetic degrees of freedom to phonon modes.  We will treat the spins
and phonons classically and in equilibrium.  With these assumptions, the
statistical mechanics of the phonons is captured by a Gaussian integral
over the associated set of displacement coordinates in the partition
function.  In such a case, the phonons can (if desired) be integrated
out to obtain an effective spin Hamiltonian which contains additional
interactions beyond the Heisenberg form.

Let us first make a few general comments regarding spin-lattice
interactions.  For a fixed, static, distortion of the lattice, we expect
modifications of the exchange interactions that are (to an excellent
approximation) linear in the displacement coordinates.  Neglecting weak
spin-orbit effects, the exchanges remain of Heisenberg form.\cite{Jia:prb05}  
Therefore the general form of the modified exchange is 
\begin{equation}
  \label{eq:modexch}
  {\mathcal H}_{\rm ex} = 
  J \sum_{\langle i j \rangle} {\bf S}_i \cdot {\bf S}_j 
  \left[ 1 - \gamma u_{ij}\right],
\end{equation}
where $u_{ij}$ is the linear combination of displacement coordinates
coupled to the pair of spins $i,j$.  Because all nearest-neighbor pairs
of pyrochlore sites are equivalent, they are all described by a single
spin-lattice constant $\gamma$.  

A na\"ive interpretation of Eq.~\eqref{eq:modexch} is that $u_{ij}$ is
proportional to the {\sl distance} between spins $i$ and $j$.  As the
distance is increased/decreased, the overlap between electronic
wavefunctions on the two sites decreases/increases, leading to a
change in the exchange coupling, proportional to this distance.  
This picture is in fact
appropriate for {\sl direct exchange}, in which there is no intervening
oxygen as in superexchange.  In the spinel chromites, the
antiferromagnetic Cr-Cr is indeed believed to arise from direct
exchange.\cite{Wickham:pr59}   More generally, the dependence of exchange 
on displacements may involve changes in the bond angles as well as
distances.  Nevertheless, from this simplistic view, one expects
$\gamma>0$ (hence the minus sign in Eq.\eqref{eq:modexch}), and $\gamma$
of order the inverse of the effective Bohr radius of the electronic
orbitals involved.  

\subsection{Bond phonon model}

In the model of Penc {\sl et al},\cite{Penc:prl04,Penc:05} the $u_{ij}$ are taken as
independent parameters, i.e. the length of each pyrochlore bond can
independently expand or contract.  This ``bond phonon'' (BP) model has
the elastic energy
\begin{equation}
  \label{eq:1}
  {\mathcal H}_{\rm ph}^{BP} = \frac{k_{BP}}{2} \sum_{\langle ij\rangle} u_{ij}^2,
\end{equation}
where $k_{BP}$ is an elastic constant.  Because each $u_{ij}$ couples only to
a single nearest-neighbor pair of spins, and in this model each $u_{ij}$
is independent, the spin-lattice interaction does not induce any
effective interactions amongst further neighbor spins.  Instead,
integrating out the $u_{ij}$ according to
\begin{equation}
  \label{eq:howtointegrate}
   e^{-\beta {\mathcal H}_{\rm eff}^{\rm BP}[\{ {\bf S}_i \}]} = \prod_{\langle
     ij\rangle} \int du_{ij}\, e^{-\beta\left( {\mathcal H}_{\rm ex}[\{
       {\bf S}_i , u_{ij}\} ] +
       {\mathcal H}_{\rm ph}^{BP}[\{ u_{ij} \}]\right)},
\end{equation}
one obtains, up to a constant, an effective spin Hamiltonian of the form
\begin{equation}
  \label{eq:2}
  {\mathcal H}_{\rm eff}^{BP} = J \sum_{\langle ij\rangle} \left[{\bf
      S}_i \cdot {\bf S}_j  - b\left({\bf S}_i \cdot {\bf S}_j
    \right)^2\right]. 
\end{equation}
Thus, in this BP model, the spin-lattice coupling induces an effective
bi-quadratic interaction or relative strength $b=\frac{\gamma^2
  J}{k_{BP}}$ between nearest-neighbor spins.  Note that this term
favors configurations in which neighboring spins are either parallel or
antiparallel, i.e. collinear configurations.  The BP model therefore
gives a simple explanation for the preference for 3:1 states in the
field range in which the classical Heisenberg model prefers
half-magnetization states.\cite{Bergman:prl05,Bergman:prb05} 

The preference for collinear spin arrangements can be understood physically in terms of the phonons as follows.
If a given pair of spins is antiferromagnetically aligned, then the
spin-lattice coupling in Eq.\eqref{eq:modexch} can be made most
negative by choosing $u_{ij}<0$, i.e. contracting the bond to enhance
the effective exchange.  Conversely, if a pair of spins is
ferromagnetically aligned, the bond can expand ($u_{ij}>0$) to weaken
the ferromagnetic exchange interaction.  In either case, the bond
energy is lowered by the same amount (because of the linear phonon
coupling) relative to the undistorted bond.

It is straightforward to see that, as claimed earlier, all the 3:1
plateau configurations remain degenerate within the BP model.  By rewriting
the exchange interaction as in Eq.~\eqref{Heisenberg}, one obtains
\begin{equation}
  \label{eq:sumsquares}
  {\mathcal H}_{\rm eff}^{BP} =  \frac{J}{2} \sum_t \left[ ({\bf S}_t
    - {\bf h})^2  - {\bf h}^2 \right]   - bJ \sum_{\langle ij\rangle}
  \left({\bf S}_i \cdot {\bf S}_j \right)^2.   
\end{equation}
For $h=2$, every 3:1 configuration minimizes the first term on each
tetrahedron as well as the second term on each link.  Hence they are
the global ground states and all degenerate.  Furthermore, even for
different values of $h$, all these states remain degenerate, since
they have the same (no longer minimal in exchange energy) ${\bf S}_t$
on each tetrahedron, and the same (minimum in energy) value of the
bi-quadratic term.  We will return to discuss the magnetization curve
and configurations away from the plateau in Sec~\ref{sec:away-from-half}.

\subsection{Einstein (site) phonon model}

The lack of splitting of the degeneracy of the 3:1 states is a
non-generic feature of the BP model.  It arises from the fact that
the bond displacements are taken to be completely independent of one
another, so that they can induce no spin correlations beyond
nearest-neighbor.  
In reality, however, this is not the case.  
To make a change in a
given bond length requires moving one or both of the atoms involved,
which will at least distort the other bonds connected to these atoms.
A more natural phonon model can be formulated in terms of the independent
displacements of each {\sl atom}, with the bond distances determined from
these atomic displacements.  If the harmonic phonon energy is taken to
be a sum of independent restoring forces for each atom, this is simply
the conventional Einstein model.  As usual, such an Einstein model
provides a crude but reasonable approximation, provided the most
important phonons are optical phonons rather than the
long-wavelength $q\approx 0$ acoustic modes.

We therefore adopt this Einstein (site rather than bond) phonon
model.  To derive an appropriate form,  let us assume a distance dependent
exchange coupling (see e.g. Ref.~\onlinecite{Jia:prb05} in a zero field context). 
It can be expanded in small atomic displacements ${\bf
  u}_i$ for each site $i$:
\begin{equation}
\begin{split} &
J_{i j} \equiv  J(|{\bf r}_i - {\bf r}_j|) \approx J({\bf R}) + ( {\bf u}_i - {\bf u}_j ) \cdot \grad J({\bf R}) + \ldots 
\\
 \approx & \left( 
1 - \gamma {\bf e}_{i j} \cdot \left( {\bf u}_i - {\bf u}_j \right)
\right) J
\,,
\end{split}
\end{equation}
where ${\bf R}$ is the vector between the pyrochlore sites $i$ and
$j$, and ${\bf e}_{i j} = {\bf R}/|{\bf R}|$ is the unit vector in the
corresponding equilibrium direction.  Comparing to
Eq.\eqref{eq:modexch}, the Einstein model has then
\begin{equation}
  \label{eq:einstein}
  u_{ij} = {\bf e}_{i j} \cdot \left( {\bf u}_i - {\bf u}_j \right), 
\end{equation}
with the elastic energy given by 
\begin{equation}
  \label{eq:Einphon}
  {\mathcal H}_{\rm ph}^{E} = \frac{k_E}{2} \sum_i |{\bf u}_i|^2.
\end{equation}

It is amusing to note that although the independent modes considered
are quite distinct in the bond and site phonon models, on the
pyrochlore lattice both formulations contain the same number of
degrees of freedom. The site displacement vectors represent $3$
degrees of freedom for each site, which for $N$  sites of the pyrochlore
lattice gives a total of $3N$ degrees of freedom. On the other hand,
there are $6$ links per tetrahedra, and one scalar bond phonon per link.
There are $N/2$ tetrahedra in the lattice, so the total number of
degrees of freedom in the bond phonon model is also $3N$.

Just as we did for the BP model, we now proceed to integrate out the
site Einstein phonons.  Because the Hamiltonian is quadratic in the
displacements, this Gaussian integration is equivalent to simply
minimizing the Hamiltonian with respect to the set of ${\bf u}_i$ and
eliminating these in favor of the spin variables.  The
optimal values of the displacements are simply
\begin{equation}
\label{eq:3}
{\bf u}_j^* = - \frac{J \gamma}{k_E} \sum_{i \in N(j)} \left( {\bf S}_i \cdot {\bf S}_j \right) {\bf e}_{i j}
\; ,
\end{equation}
where $N(j)$ denotes the set of nearest neighbors of site $j$.  At this
point, it is evident already that the presence of lattice distortions is
tied to frustration.  If all exchanges could be satisfied equally, i.e.
$\left\langle {\bf S}_i \cdot {\bf S}_j \right\rangle = const$, then the
distortion vanishes: $\langle {\bf u}_j^*\rangle \sim \sum_{i \in N(j)}
{\bf e}_{i j} = 0$.  A distorted lattice is thus induced only in
frustrated states, and for instance no distortion is expected at large
fields where the spins are fully and uniformly polarized.  If a
distortion is still observed in a fully polarized state, it cannot arise
from this mechanism, and might have no connection with the magnetic
order of the system.

Substituting back Eq.\eqref{eq:3} in the Hamiltonian, we obtain the
effective Hamiltonian for the (site) Einstein (E) model:
\begin{equation}\label{H_dist}
{\mathcal H}_{\rm eff}^{E} =
J \sum_{\langle i j \rangle} {\bf S}_i \cdot {\bf S}_j 
 - \frac{k_E}{2} \sum_j \left|{\bf u}_j^*\right|^2
\; ,
\end{equation}
where, as above, the Zeeman interaction with the external field has been
dropped for brevity.  The form in Eq.\eqref{H_dist} is actually quite
convenient for analysis, but we first write out the induced interactions
explicitly for comparison with the BP model.  One finds
\begin{eqnarray}
  \label{eq:4}
  {\mathcal H}_{\rm eff}^{E} & = & J \sum_{\langle ij\rangle} \left[{\bf
      S}_i \cdot {\bf S}_j  - b\left({\bf S}_i \cdot {\bf S}_j
    \right)^2\right] \nonumber \\
  & - & J\frac{b'}{2} \sum_{j \neq k\in N(i)}  ({\bf S}_i \cdot {\bf S}_j)({\bf S}_i
  \cdot {\bf S}_k) {\bf e}_{ij} \cdot {\bf e}_{ik}, \nonumber \\
\end{eqnarray}
with $b=\frac{J\gamma^2}{k_E}$ as before.  Here we have defined a
separate parameter $b'$ which, according to the strict development above, is not independent ($b'=b$).
However, it
distinguishes the additional terms in the effective spin Hamiltonian
which are not present in the BP model.  We propose to view $b'$ as a
separate phenomenological parameter to describe the phonon modes.  In
particular, taking $0<b'<b$ corresponds to ``softening'' the bond
phonons somewhat, and interpolating between the Einstein model and the
BP model.

Because the same biquadratic term is present as in the BP model, we
may expect that, as in that case, collinear states are favored.
Indeed, when the additional terms are weak, $b' \ll b$, they may be
viewed as weak perturbations.  Since the $b$ term already selects
collinear 3:1 configurations, the $b'>0$ term can then only select a
ground state within this set of states, and has little effect on the
stability of the plateau. We show below that $b'>0$ selects a
particular collinear classical spin configuration depicted in Fig.4,
in the manifold of 3:1 states.  Our analysis shows that for the
range of parameters $b'(<\frac{b}{2})$ the plateau retains a finite
width, and the same particular 3:1 configuration prevails.

\begin{figure}[hbt]
	\centering
	\subfigure[Site adjoining the tetrahedra is the only minority site.]{
	\label{fig:conf1}
		\includegraphics[width=1.5in]{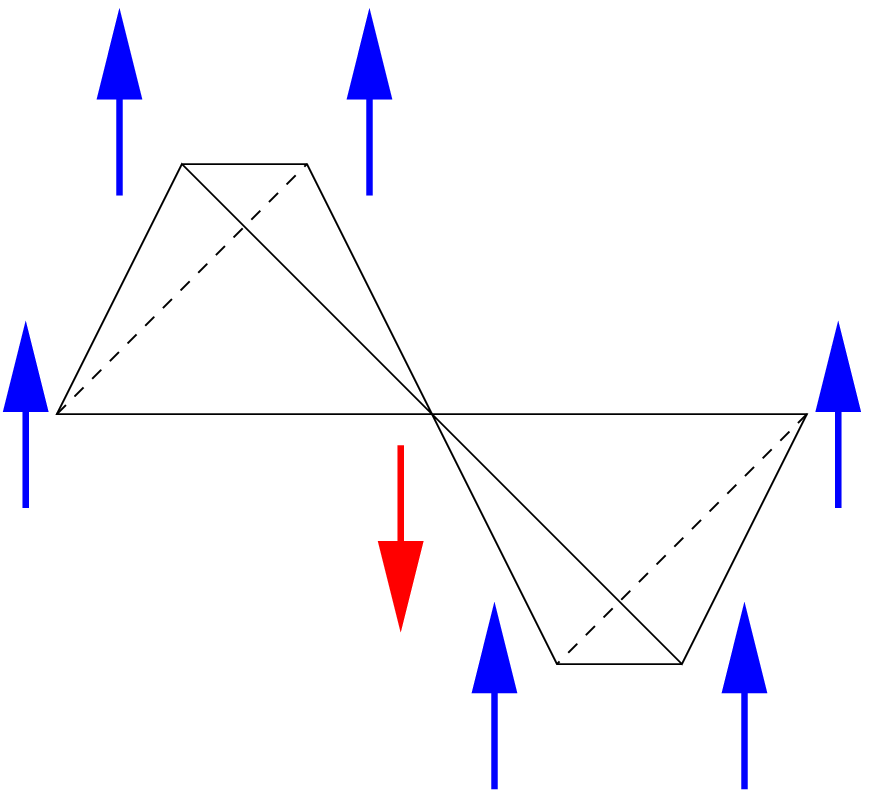}}
	\subfigure[Two minority site connected by two parallel links.]{
	\label{fig:conf2}		
		\includegraphics[width=1.5in]{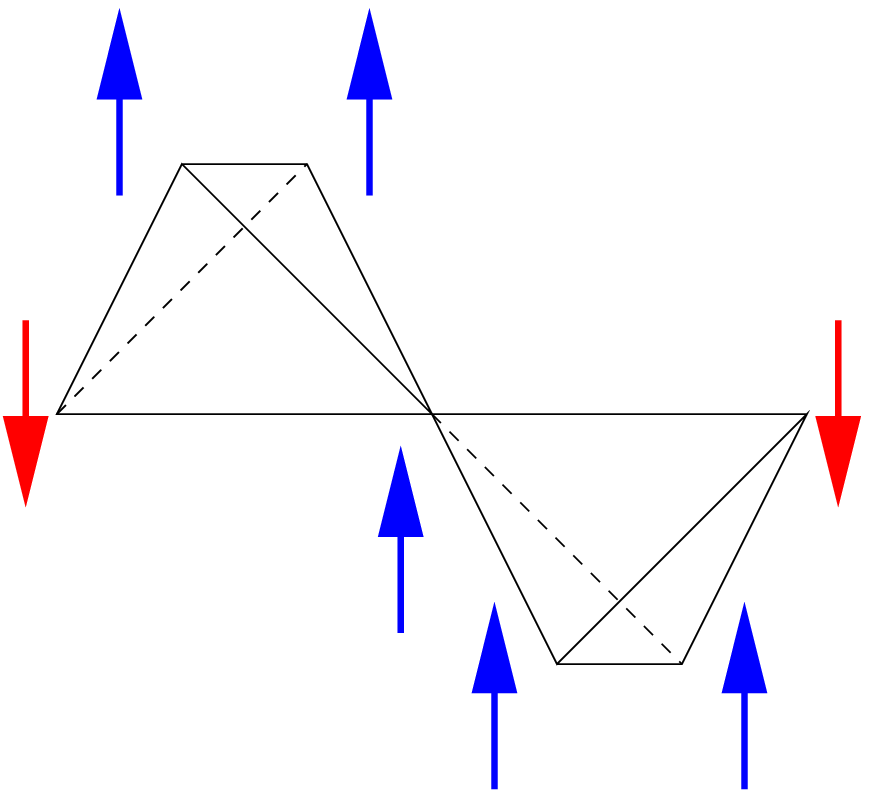}}
  \hspace{2.0in}	
  \subfigure[Two minority site connected by two links bending.]{
	\label{fig:conf3}
		\includegraphics[width=1.5in]{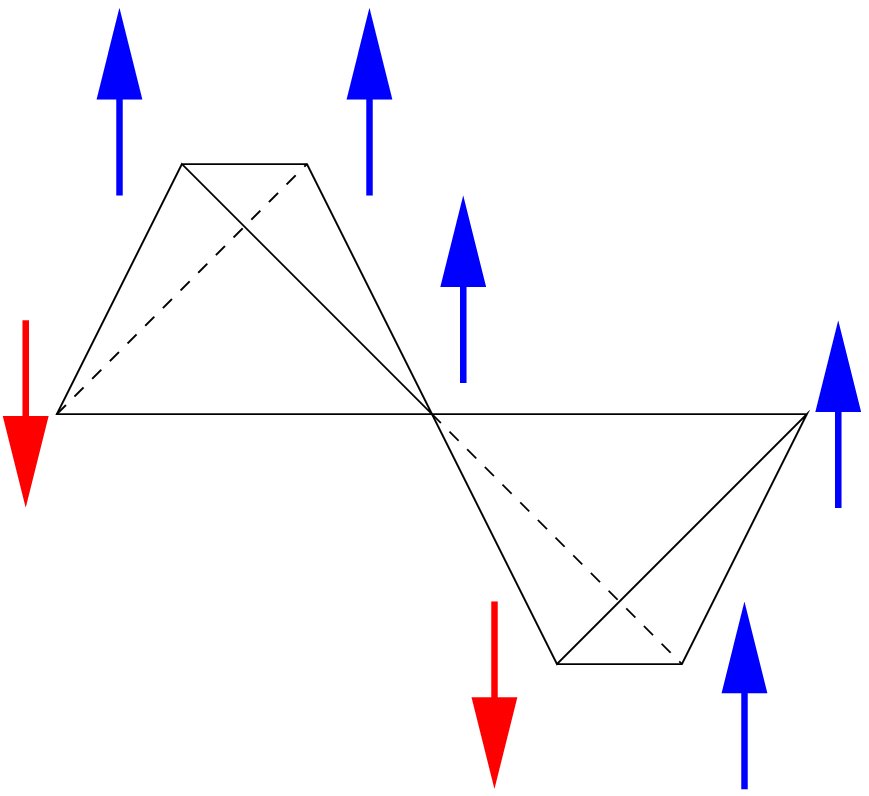}}
  \subfigure[Distortion of the bending configuration.]{
	\label{fig:forces}
		\includegraphics[width=1.5in]{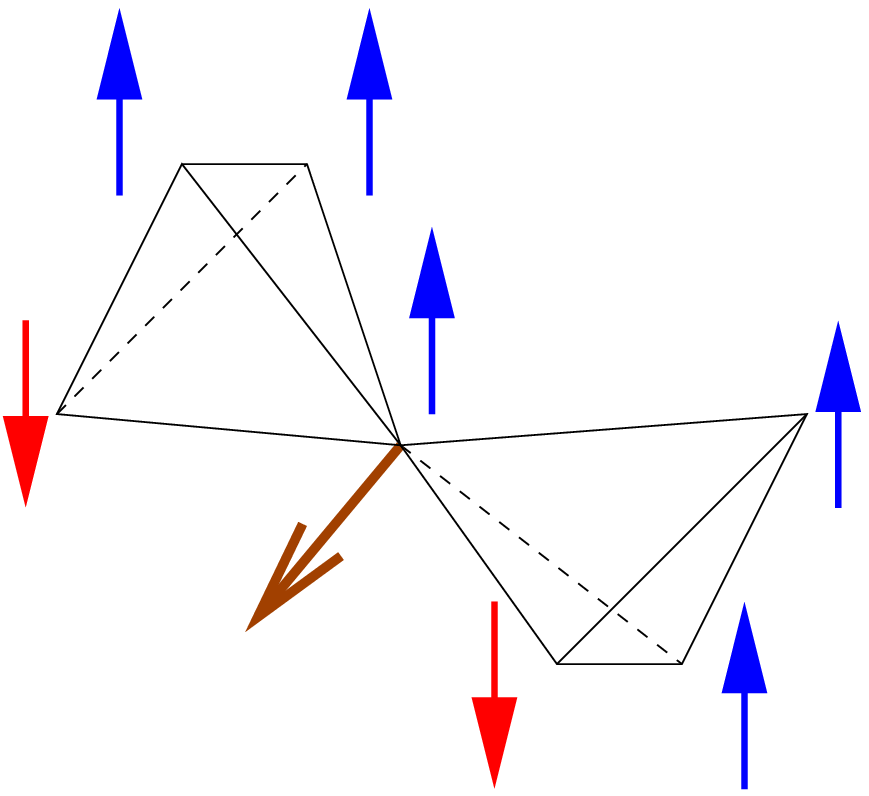}}
	\caption{(Color online) The three generic configurations of minority sites (red arrow) on two adjacent tetrahedra.}
	\label{fig:configs}
\end{figure}

\begin{figure}
	\centering
		\includegraphics[width=2.5in]{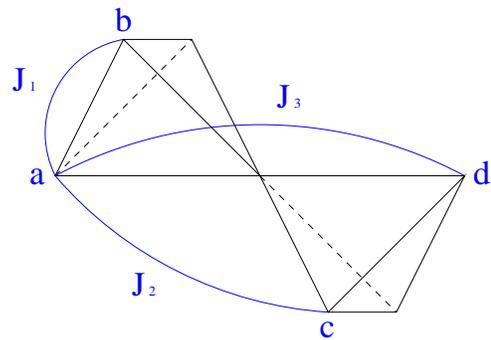}
	\caption{(Color Online) Further neighbor interactions in the
          pyrochlore lattice. The nearest neighbor interaction $J_1$ is
          between site $a$ and site $b$, the next nearest neighbor
          interaction $J_2$ is between site $a$ and site $c$, and finally
          the next next nearest neighbor interaction $J_3$ is between
          site $a$ and site $d$.} 
	\label{fig:furtherNeighbors}
\end{figure}

To get some
intuition for how additional degeneracy-breaking is induced by the $b'$
terms, assume a collinear state, ${\bf S}_i =  {\bf\hat{z}} \sigma_i$,
with $\sigma_i=\pm 1$ Ising spins satisfying the 3:1 constraint.  The
first line in Eq.\eqref{eq:4} becomes constant, and the
four-spin product $({\bf S}_i \cdot {\bf S}_j)({\bf S}_i \cdot {\bf S}_k) = 
\sigma_j \sigma_k$ becomes an effective two-spin interaction.  Then the
Hamiltonian within the 3:1 manifold takes the form of Ising exchange
terms:  
\begin{eqnarray}
  \label{eq:5}
   {\mathcal H}_{\rm 3:1}^{E} & = &  J^{\rm eff}_2 \sum_{\langle\langle
     ij\rangle\rangle} \sigma_i \sigma_j + J_3^{\rm eff}
   \sum_{\langle\langle\langle kl \rangle\rangle\rangle}
   \sigma_k\sigma_l ,
\end{eqnarray}
where we have dropped some terms which are constant in the 3:1 manifold.
The parameters $J_2^{\rm eff}=J b'/4$ and $J_3^{\rm eff}=J b'/2$ are the effective Ising
exchanges between second-neighbor sites (connected by two bent links as
in Fig.\ref{fig:conf3}) and third-neighbor sites (connected by two
parallel links as in Fig.\ref{fig:conf2}, and explicitly shown in Fig.~\ref{fig:furtherNeighbors} ).  Note that $J_3^{\rm eff}$ is twice as
large as $J_2^{\rm eff}$, so the third neighbor interactions tend to be favored
over second neighbor interactions in the determination of the magnetic state.

Clearly these effective interactions break the degeneracy of the 3:1
manifold.  To understand this breaking more physically, and thereby
derive the selected ordered state on the plateau, we return to the
formulation of Eq.\eqref{H_dist}.  From Eq.\eqref{H_dist} it is readily apparent that the
larger the induced lattice displacement, the lower the energy.
For $b\ll 1$, we may assume half-magnetization on each tetrahedron in
the plateau region, and then consider the effect of the lattice
displacement as a perturbation.  


Let us begin by first assuming the 3:1 constraint on each
tetrahedron.  We then wish to understand which configurations of
minority spins maximize the displacements.  Consider 2 tetrahedra
adjoined by a site $j$. There are 3 generic configurations of the
positions of the minority sites, depicted in Fig.~\ref{fig:configs}.  
A {\em fixed} fraction ($1/4$) of all configurations are necessarily those in
Fig.\ref{fig:conf1}, therefore 
the energy of this configuration is irrelevant to the splitting.   
Of the remaining two configurations, it
is simple to understand that a nonzero ${\bf u}_j^* $ can only arise in
the configuration Fig.~\ref{fig:conf3}, because it is the only
configuration with asymmetry about the site $j$.   Therefore, we wish to
maximize the number of configurations of this type.

\begin{figure}[hbt]
	\centering
		\includegraphics[width=4.0in]{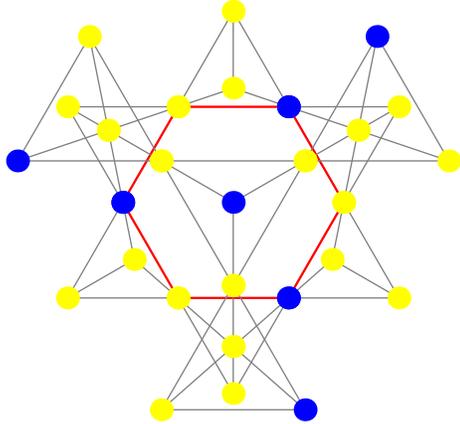}
	\caption{(Color online) Spin configuration of the R state. Majority sites are colored yellow, minority sites are colored blue. The flippable plaquette in this unit cell of the R-state configuration is highlighted in red. }
	\label{fig:R_Config}
\end{figure}

We can describe this favored configuration with a ``bending rule''.
For every pair of adjacent tetrahedra adjoined by a majority site, 
mark the links connecting between the two minority sites. These marked links form paths on the lattice, 
connecting all the minority sites. It is energetically preferable for these these paths
to bend, rather than continue on a straight line.   
Clearly, the maximum
number of such bent paths is obtained if {\sl all} paths are bent,
i.e. all minority sites are in the configurations in
Fig.\ref{fig:conf3}.  This can indeed be achieved.  By careful
enumeration of all configurations ( see appendix~\ref{app:bending} for details), it can be 
shown that there is a unique (up to lattice symmetries) configuration 
which satisfies this ``bending rule'' on a ``pyrochlore cell'' -- 
a volume containing 4 hexagonal plaquettes (see Fig.~\ref{fig:R_Config}).  
If this configuration is required on all such cells, the ground state is 
completely specified. 
This is precisely the ``R'' state obtained in Ref.~\onlinecite{Bergman:prl05}
 in a very different quantum dimer model analysis.   The ``R'' state has space group $P4_332$
 and may be thought of in terms of filling the pyrochlore lattice with a fraction $1/4$ of hexagonal plaquettes with alternating up/down spins and a fraction $3/4$ of plaquettes with one down spin and
 five up spins.

Because the argument above proves the R-state is the best possible
collinear state, any alternative ground state must be non-collinear.
Since non-collinear states cannot exhibit a plateau, proving that the
R-state is the global minimum energy state is equivalent to proving
the existence of a plateau.  However, it is important to emphasize
that the above argument assumes the 3:1 configurations (actually it
can be made equally rigorous assuming only collinearity).  For $b' \ll
b$, this assumption is valid, and the above argument becomes
controlled.  For the generic situation with $b' \sim b$, the effective
Hamiltonian is actually ``frustrated'' in the following sense.
Because the Einstein phonon displacement resides on a pyrochlore site
and is related via Eq.\eqref{eq:3} to spins on the two neighboring
tetrahedra, the natural unit for the effective Hamiltonian is no
longer a bond but such a pair of adjacent tetrahedra.  One may rewrite
the Hamiltonian as a sum over such pairs, parametrized by the
pyrochlore site $j$ they share: ${\mathcal H} = \sum_j {\mathcal H}_j$
, with
\begin{equation}
\begin{split}
\label{H_coup_ave}
{\mathcal H}_j = &
\frac{J}{8} \left[ \left( {\bf S}_{t_1} - {\bf h} \right)^2  - h^2 \right]
+  \frac{J}{8} \left[ \left( {\bf S}_{t_2} - {\bf h} \right)^2  - h^2 \right] 
\\ &
- J \frac{b}{4} \sum_{\langle ik \rangle \in t_{1,2}} \left(
{\bf S}_i \cdot {\bf S}_k  
\right)^2
\\ &
-  J \frac{b'}{2} \sum_{i \neq k\in N(j)}  ({\bf S}_i \cdot {\bf S}_j)({\bf S}_j
  \cdot {\bf S}_k) {\bf e}_{ji} \cdot {\bf e}_{jk}
\; . 
\end{split}
\end{equation}

The extra factors of $\frac{1}{4}$ above relative to
Eq.\eqref{Heisenberg} in the nearest neighbor terms are due to the fact that every tetrahedron is
shared by $4$ pyrochlore sites.  One can show that for $b,b'>0$, for a
non-zero window of fields $h$ in the neighborhood of $h=2$ (the plateau
region), ${\mathcal H}_j$ is minimized,
when the pair of adjacent tetrahedra sharing site $j$ is in the
``bending'' configuration pictured in Fig.\ref{fig:conf3}.  Clearly,
however, this condition cannot be {\sl simultaneously} satisfied on
every tetrahedral pair, because some of the tetrahedron pairs must
have a minority site adjoining them.  Thus the Einstein site phonon
model exhibits ``tetrahedral-pair frustration''.  The R-state argued
for above resolves this frustration in a natural way, by minimizing
the energy on a maximal fraction of tetrahedral pairs (which is
$3/4$).  Because such a relatively large fraction of tetrahedron pair
units are ``satisfied'', it appears plausible that the R-state is
indeed the global ground state.

We have searched numerically to check for the alternative possibility,
that a lesser fraction (perhaps zero) of units are satisfied, but that
the energy of the dis-satisfied units is sufficiently better as to
lower the total energy.  In order to take the ``tetrahedral-pair
frustration'' into account, we must go beyond the above
single-tetrahedron analysis for the BP model.  In particular, we must
consider units larger than a single tetrahedron, and also larger than
a single tetrahedron-pair unit: since these units are frustrated, they
cannot be simultaneously satisfied at most fields.  Instead, we
consider a cluster of five tetrahedra consisting of a central
tetrahedron and its four surrounding neighbors.  This is the smallest
collection of tetrahedra for the 3:1 states in which a pair in the 
unsatisfied configuration of
Fig.~\ref{fig:conf1} \emph{must} appear.
The Hamiltonian can be written as a sum over the up pointing
tetrahedra ${\mathcal H} = \sum_{t \in \u} {\mathcal H}_t$, $t$ being
the central tetrahedron in each cluster. The down pointing tetrahedra
are counted in 4 different clusters in this scheme, so we account for
this by writing the cluster Hamiltonian as
\begin{equation}
\label{H_cluster_ave}
\begin{split}
{\mathcal H}_t = &
\frac{J}{2} \left[ \left( {\bf S}_{t_0} - {\bf h} \right)^2  - h^2 \right]
+ \frac{J}{8} \left[ \left( {\bf S}_{t_1} - {\bf h} \right)^2  - h^2 \right] 
\\ &
+ \frac{J}{8} \left[ \left( {\bf S}_{t_2} - {\bf h} \right)^2  - h^2 \right] 
+ \frac{J}{8} \left[ \left( {\bf S}_{t_3} - {\bf h} \right)^2  - h^2 \right]
\\ &
+ \frac{J}{8} \left[ \left( {\bf S}_{t_4} - {\bf h} \right)^2  - h^2 \right] 
- J \frac{b}{4} \sum_{\langle i k \rangle \in t_{0,1,2,3,4}} \left(
{\bf S}_i \cdot {\bf S}_k  
\right)^2
\\ &
-  J \frac{b'}{2} \sum_{j=1}^4 \sum_{i \neq k\in N(j)}  ({\bf S}_i \cdot {\bf S}_j)({\bf S}_j
  \cdot {\bf S}_k) {\bf e}_{ji} \cdot {\bf e}_{jk}
\; ,
\end{split}
\end{equation}
where $t_0$ is the middle tetrahedron, $t_{1..4}$ are the other four tetrahedra, and the sites $j=1...4$
are the 4 corners of the tetrahedron $t_0$.

The conclusion from our numerical study, is that for $0<b'\lesssim
b/2$, the minimum of ${\mathcal H}_t$ above is indeed a 3:1
configuration comprising a corresponding section of the R-state.
Hence, because such a configuration can be simultaneously realized on
every such unit, in this parameter range, the spin-lattice coupling
indeed stabilizes a state with the R-state symmetry.  The width of the
corresponding plateau is discussed in the following section.

\section{Away from half-polarization}
\label{sec:away-from-half}

In this section, we explore the properties away from the magnetization plateau.

\subsection{BP Model}
\label{sec:bp-model}

Let us first review the findings of Penc {\sl et al}\cite{Penc:prl04}
in the BP model.  The basic results can be understood by simple
considerations on a single tetrahedron.  Such a simplification is
satisfactory in this case because the Hamiltonian can be written as a
sum of such terms on each tetrahedron, and they can be simultaneously
satisfied.  Thus the BP model does not suffer from ``tetrahedral-pair
frustration''.

For magnetization greater than half polarization, the ground state has
a 3:1 configuration, with 3 majority spins and 1 minority spin on each
tetrahedron.  However, they are not collinear, except on the plateau
and at saturation.  These vary in such a way that the net spin per
tetrahedron is increased from $2$, by {\sl smoothly} rotating the
minority spin from down to up.  Because the collinear plateau state
can be smoothly deformed into the states above the plateau, there is a
{\sl continuous} transition at the upper plateau
edge.\cite{Penc:prl04}
All these states above the plateau share the same degeneracy as
the plateau states: the location of the minority spins is not 
determined in the BP model.  

On the other hand, a state with magnetization per tetrahedron of less
than $2$ cannot be achieved with a 3:1 configuration.  Thus, for
fields below the plateau, the structure of the configurations {\sl
  must} change.  Instead, over most of this region of phase space, the
spin configuration assumes a 2:2 form.  This implies a discontinuous
change of spins, and gives rise to a {\sl first order}
transition.\cite{Penc:prl04} Like the 3:1 states, these 2:2 states are
highly degenerate, due to the many equivalent manners in which each of
the 2 spin polarizations may be arranged.  Both above and below the
plateau, the low temperature phases break rotational symmetry about
the field axis, but with no net moment in the $x-y$ plane.

\subsection{Einstein model}
\label{sec:einstein-model}

\subsubsection{Magnetization}

The BP model captures rather well the broad behavior of the low-temperature
magnetization curve, $M(H)$, in the chromites, where it has been
observed.  The only qualitative exception is the observation of a small
feature at $H\approx 37T$ for \hgaf, in the field range between the plateau and
saturation, which has been suggested as an additional phase transition.\cite{Ueda:prb06}

\begin{figure}
	\centering
		\includegraphics[width=3.5in]{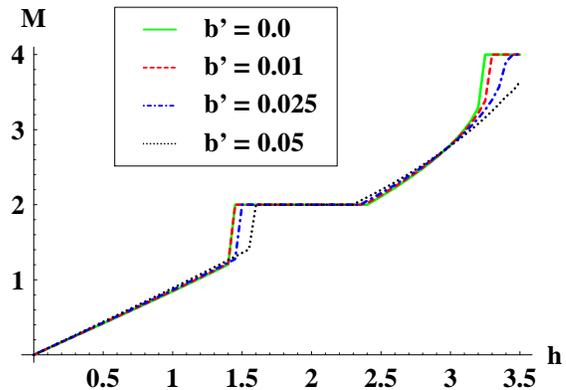}
                \caption{(Color online) Magnetization curves for various values of $b'$ with $b=0.1$
                fixed, where $b$ and $b'$ are
                  given below Eq.\eqref{H_coup_ave}. The plateau width decreases with growing $b'$. }
	\label{fig:Mag_curve}
\end{figure}

The magnetization curves we obtained for the Einstein model ($b' < b$)
are similar to those for the BP model over most field values.
Studying the magnetization curve for the Einstein model, however, is much more
involved than the above single-tetrahedron analysis for the BP model,
due to the ``tetrahedral-pair frustration'', explained in the previous
section.  Following the above treatment, we consider a
collection of five tetrahedra.  On this cluster, we numerically
minimize the energy \eqref{H_cluster_ave} for each field, and
determine the zero-temperature $M(H)$ curve for given values of
$b,b'$. The magnetization curves for one value of $b=0.1$ and various
values of $b'$ are plotted in Fig.~\ref{fig:Mag_curve}.

In the magnetization curve shown in Fig.~\ref{fig:Mag_curve} an abrupt
jump is observed going from the linear regime into the half
magnetization plateau region, in accordance with the results of Penc
et al.\cite{Penc:prl04}  Given the 2:2 spin configuration at zero
magnetic field, and the 3:1 configuration in the plateau region, we
can contemplate a sharp transition onto the plateau between 2:2 and
3:1 configurations. Such a transition would have to be first order,
as the symmetry groups are not Landau-Ginzburg compatible for a second
order transition (one is not the subgroup of the other).

\subsubsection{Phases}

As we have seen above, the differences in the magnetization curves of
the Einstein and BP models are minimal.  The magnetization, as a
purely thermodynamic quantity, is only weakly sensitive to the
detailed correlations between spins beyond one or two lattice
spacings.  A better test of the differences between the models is to
compare their phase diagrams.  In the BP model, as described above,
their are four ``phases'', in which the local structures on each
tetrahedra are distinct: at low fields, a 2:2 structure or a
``splayed'' structure; on the plateau, a collinear 3:1 structure;
above the plateau, a non-collinear 3:1 structure, and at high fields,
the ferromagnetic fully saturated configuration.\cite{Penc:prl04} We
have used quotation marks around the word ``phases'' because all but
the ferromagnetic configuration exhibit an unphysical macroscopic
degeneracy not related to symmetry.

A full determination of the phase diagram in the Einstein model is
beyond the scope of this paper.  However, we will outline those
features which are similar and those which are clearly distinguishable
from the BP model.  At zero field, it can be shown that the ground
states of the Einstein model are far less degenerate for all $b'>0$
than those of the BP model.  To see this, we use the representation in
Eq.\eqref{H_coup_ave}, and consider the minimum energy configuration
on a single tetrahedral-pair with $h=0$.  Simple analysis shows that
this minimum energy occurs for collinear states with a 2:2 ratio of
``up'' and ``down'' spins on each of the two tetrahedra, satisfying
the ``bending rule'' if links are drawn between the spins aligned with
the central one.  Because the field $h=0$, the spin axis is arbitrary.
Unlike in the plateau region, such 2:2 states are {\sl unfrustrated}:
every tetrahedral pair can be chosen to have such a configuration.  In
fact, these states are still highly degenerate.  They correspond to
``ice-rules'' states, with the additional constraint of the ``bending
rule''.  Though we do not have an analytical calculation of the number
of such states, we have performed a numerical enumeration of them on
various finite size clusters (see
Table~\ref{tab:BendingIceConfigurations}).  Evidently, the number of
such ``bending ice'' states grows rapidly with system size.  It is
likely that these states are macroscopically degenerate.
Nevertheless, this set of states is much less degenerate than the zero
field ground states in the BP model, which are simply {\sl all} the
2:2 ``ice'' states, without the bending rule imposed (see
Table~\ref{tab:IceConfigurations}).  We will comment
upon the physical consequences of this degeneracy in
Sec.~\ref{sec:discuss}.

\begin{table}
\begin{tabular}{|c|c|}
\hline \hline
Number of unit cells  & Number of bending ice configurations  \\
\hline 
$2 \times 2 \times 2 = 8$  &  $12$ \\
$2 \times 2 \times 4 = 16$ &  $36$ \\
$2 \times 4 \times 4 = 32$ &  $82$ \\
$4 \times 4 \times 4 = 64$ &  $216$
\\
\hline
\end{tabular}	
	\caption{Bending Ice configurations. Dimensions of pyrochlore cluster indicated. Periodic boundary conditions were used.}
	\label{tab:BendingIceConfigurations}
\end{table}

\begin{table}
\begin{tabular}{|c|c|}
\hline \hline
Number of unit cells  & Number of ice configurations  \\
\hline 
$2 \times 2 \times 1 = 4$  &  $78$ \\
$3 \times 2 \times 1 = 6$ &  $534$ \\
$2 \times 2 \times 2 = 8$ &  $2970$ \\
$3 \times 3 \times 1 = 9$ &  $7974$ \\
$5 \times 2 \times 1 = 10$ &  $28326$ \\
$3 \times 2 \times 2 = 12$ &  $87684$ \\
\hline
\end{tabular}	
	\caption{Ice configurations. Dimensions of pyrochlore cluster indicated. Periodic boundary conditions were used.}
	\label{tab:IceConfigurations}
\end{table}

At small non-zero fields, we expect the same ``bending rule'' states
to remain approximate ground states, with the axis of the two spin
orientations ``flopped'' into a fixed one at a small angle
(proportional to $h/J$) away from the $x-y$ plane in spin space.  Thus
in this region there is a broken XY symmetry around the spin-rotation
axis. Indeed numerical minimization of a tetrahedron pair shows that
for finite weak magnetic field the nearly collinear 2:2 bending state
persists.  For intermediate fields half-way between zero field and the
plateau, we do not have definitive results.  As in the BP model, there
is a first order transition separating the low-field region from the
plateau.

Like in the BP model, at fields just above the plateau, the 3:1
structure deforms smoothly into a state with larger than half
polarization by small rotations of the spins.  Because this
deformation is smooth, we expect that the space group symmetry of the
``canted ferrimagnetic'' state just above the plateau will be {\sl at
  least as low} as the $P4_332$ symmetry of the R-state.  This should
be observable in neutron scattering as the persistence of magnetic
scattering peaks -- present in $P4_332$ but not the $Fd\overline{3}m$
space group of the ideal spinel -- in the field region above the
plateau.  This is a distinct prediction of the Einstein model.

In addition to this persistent discrete symmetry breaking, this region also
exhibits XY long range order of the spin components perpendicular to the
field.  The nature of this XY order is not apparent from simple
arguments.  Classically, it can be analyzed for the Einstein model by
assuming small deformations of the spins,
\begin{equation}
  \label{eq:7}
  {\bf S}_i = \left( {\rm Re}\psi_i, {\rm Im}\psi_i, \sigma_i
    \sqrt{1-|\psi_i|^2}\right), 
\end{equation}
where $\sigma_i=\pm 1$ is the Ising magnetization of spin $i$ in the
R-state, and $\psi_i$ is the transverse spin represented as a complex
vector.  By inserting this into Eq.\eqref{eq:4} and expanding to
quadratic order, one obtains a quadratic form in $\psi_i$.  This can be
diagonalized using Bloch's theorem to obtain a set of 16 (one per site
of the R-state unit cell) ``bands''.  The first vanishing eigenvalue(s)
of this spectrum on increasing field signals the upper edge of the
plateau.  The eigenfunction (wavevector(s) and wavefunctions) is the
order parameter of the XY magnetism in the canted ferrimagnet.  

\begin{figure}
	\centering
		\includegraphics[width=2.0in]{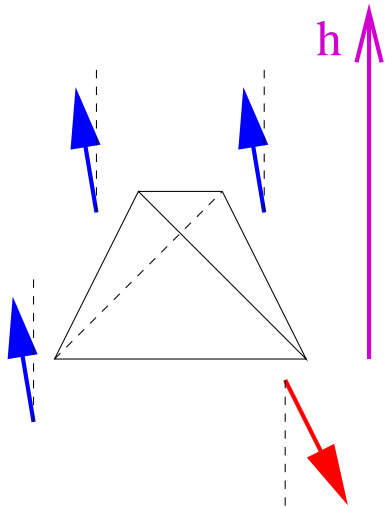}
	\caption{(Color online) Spin configuration on one tetrahedron for magnetic
          fields just above the half magnetization plateau.}
	\label{fig:off_plateau}
\end{figure}

We have carried out this calculation for $b=0.1$ and
$b'=0.0,0.01,0.025,0.05$ (the same values for which we plotted the
magnetization curve in Fig.~\ref{fig:Mag_curve}), and a range of
magnetic fields sweeping through the transition off the plateau. Our
findings are that the excitation minimum is at ${\bf k} = 0$ (the
$\Gamma$ point), with an eigenfunction which retains all the point
group symmetries of the R-state.  The resulting non-vanishing XY
components of the spins just above the plateau are equal on all
minority sites and equal on all majority sites.  However, their
direction is opposite on the two types of sites, and the magnitude of
the transverse component on the minority sites is always larger than
that of the majority sites, by a factor which is very close but not
precisely equal to $3$. This indicates that there is a {\sl
  non-vanishing total XY magnetization} -- transverse to the field
axis.  It is very small, but non-zero.  The magnetic state is
demonstrated in Fig.~\ref{fig:off_plateau}.  Physically, this would be
visible in a single-domain sample as a spontaneously deviation of the
magnetization axis from that of the applied field upon leaving the
plateau.  

The reduced space group symmetry (relative to $Fd\overline{3}m$) of
this ``canted ferrimagnetic'' phase implies that there {\sl must} be
at least one phase transition between this state and the fully
saturated ferromagnetic state, even at $T>0$.  Moreover, if there is
only a single phase transition, a standard Landau analysis -- see
Appendix~\ref{app:LG_theory} -- predicts it cannot be continuous.
This is not the case in the BP model, where for small $b$ one observes
a single continuous transition into the ferromagnetic state.  There
are two possibilities: a first order transition from the canted
ferrimagnet to the fully polarized state, or an {\sl intermediate
  phase transition} between the $P4_332$ (R-state structure) canted
ferrimagnet and a partially-polarized canted ferromagnet with higher
space group symmetry.  The latter transition is a possible explanation
of the observed magnetization feature at $H\approx 37T$ in
\hgaf.\cite{Ueda:prb06} Some theoretical aspects of the magnetic
behavior near saturation fields have been investigated in
Ref.[\onlinecite{Derzkho:prb05}].

\section{Further neighbor interactions}
\label{sec:J23}

In this section we consider how the degeneracy on the plateau might be
broken by further neighbor interactions.  We take as our model the BP
Hamiltonian (Eq.\eqref{eq:2}) {\sl plus} additional second and third
neighbor exchange interactions:
\begin{equation}
\label{eq:6}
{\mathcal H} = {\mathcal H}_{\rm eff}^{BP}+
J_2 \sum_{\langle \langle i j \rangle \rangle} {\bf S}_i \cdot {\bf S}_j +
J_3 \sum_{\langle \langle \langle k \ell \rangle \rangle \rangle} {\bf
  S}_k \cdot {\bf S}_\ell \; .
\end{equation}
The corresponding pairs of sites were indicated 
in Fig.\ref{fig:configs}.  

Since the ground states of the BP model are exactly the set of 3:1
collinear states, we may treat the additional small $J_2$ and $J_3$
exchange couplings in Eq.\eqref{eq:6} as perturbations.  To leading
order, this amounts to simply replacing ${\bf S}_i = \sigma_i
{\bf\hat{z}}$ with Ising spins $\sigma_i=\pm 1$ satisfying the 3:1
constraint.  Doing so, one obtains {\sl the same effective Ising
  Hamiltonian} as we found in the spin-lattice model, Eq.\eqref{eq:5},
but with the effective Ising couplings replaced by the physical second
and third neighbor exchanges, $J_{2/3}^{\rm eff}=J_{2/3}$! Thus it is
immediately obvious that if $J_2$ and $J_3$ are chosen in a way to
match those of the Einstein model (i.e. $J_3=2J_2$), the R-state will
again be favored.

What state is selected by more general exchange interactions? Consider
two adjacent tetrahedra. If the minority spin is at the site joining
the two, this is the only configuration the constraint will allow
(Fig.~\ref{fig:conf1}).  Otherwise, two minority sites must be
present, in any configuration.  This pair can either be next-nearest
neighbors (Fig.~\ref{fig:conf3}) or $3^{rd}$-nearest neighbors
(Fig.~\ref{fig:conf2}).  The latter is preferred when $J_3 > J_2$, and
otherwise the former tends to be preferred.  This can easily be seen
by calculating the total Ising energy of both configurations. The
Ising Hamiltonian is satisfied by having \emph{all} pairs of
tetrahedra in the preferred (``bent'') generic configuration of the
two. This is exactly the same ``bending'' rule found in the previous
section, arising from coupling to lattice distortions. Therefore, the
R-state will be the ground state if and only if $J_3>J_2$.  If the
opposite relation holds, $J_3<J_2$, ``unbent'' configurations are
instead favored.  This leads to a rather different state with
$R\bar{3}m$ symmetry.  In this state, the magnetic unit cell is {\sl
  not} enlarged relative to the non-magnetic one.  Instead, all
equivalent (i.e. ``up'' or ``down'') tetrahedra have {\sl the same}
specific 3:1 configuration.  This state is thus directly analogous to
the ``uud'' state expected for a kagome antiferromagnet in a field.
It is only 4-fold degenerate (compared to the 8-fold degeneracy of the
R-state).  

One may also study the further-neighbor exchange model at higher and
lower magnetic fields.  By straightforward application of the same
methods used in Sec.\ref{sec:einstein-model}, we find that for
$J_3>J_2$, the further-neighbor exchange interactions favor exactly
the same canted ferrimagnetic state as the Einstein model just {\sl
  above} the magnetization plateau (see Fig.~\ref{fig:off_plateau}).  

At zero field, the situation is more interesting.
Consider a pair of adjacent tetrahedra, as in Fig.~\ref{fig:furtherNeighbors}.
This is the smallest cluster of tetrahedra that includes spins interacting via 
further neighbor exchange. 
If the coupling $b$ is assumed small relative to $J_2, J_3$, 
and $J_2>J_3$ then the ground state is collinear, preferring 
a configuration of parallel spins on a straight line,
as opposed to a ``bending ice'' configuration.
The $J_2<J_3$ case is however more complicated, and the outcome unclear
from our limited analysis of a single pair of tetrahedra.
Setting $b=0$, and assuming the magnetization
on each tetrahedron vanishes, if $J_2 = J_3$ we find the further neighbor interactions
sum to a constant. As a result we can subtract $J_2$ from $J_3$ and deal only with the 
third neighbor interaction terms, with the coefficient $J_3 - J_2 > 0$. Minimizing the third 
neighbor interaction terms on this single pair of tetrahedra we find a non-collinear state with a minimum energy of 
$-\frac{7}{3} \left( J_3 - J_2 \right)$. The ``bending ice'' collinear state gives an energy
of $-\left( J_3 - J_2 \right)$, significantly higher.
In the non-collinear state, the four spins on each of the two tetrahedra point in four
different directions defining a tetrahedron {\sl in spin space}, i.e.
with all neighboring pairs of spins at $\approx 109^\circ$ angles to
one another. However, we  \emph{cannot} tile the entire lattice with this
configuration on each pair of tetrahedra.
Given the large energy difference
between the collinear bending state and the non-collinear state on the single pair, it is reasonable to expect 
that some other non-collinear state realizes the energy minimum.
Of course, as $b$ is increased, collinear
states must be favored even at zero field.  For sufficiently large
$b$, the ground states for $J_3>J_2$ are the identical ``bending ice''
configurations found in the Einstein model.

Other possible interactions in the chromites include dipolar
interactions. Spin ice compounds, for example, often have appreciable
dipole-dipole interactions\cite{Bramwell:01} that cannot be ignored.
Using the estimate of the magnetic moment of Cr$^{+3}$ ions in \cdaf,
$\mu \simeq 3.7 \mu_B$ we estimate the dipole-dipole coupling strength
to be $ D = \frac{\mu_0}{4 \pi} \frac{\mu^2}{r_{nn}^3} \simeq 0.3 K $
where $r_{nn} = 8.567 A$ is the nearest neighbor distance in the \cdaf
\, pyrochlore structure.  With the experimental observation of the
magnetic ordering temperature at a few degrees Kelvin, it is evident
that the energy scale of the dominant degeneracy breaking mechanism,
regardless of its details, is at least an order of magnitude larger than
that of dipolar interactions. For this reason we believe that the
dipolar interactions play no significant role in the physics of the
chromite spinels, and can be safely neglected.

\section{Discussion}
\label{sec:discuss}

In this paper we have studied classical mechanisms of degeneracy
breaking in pyrochlore anti-ferromagnets and how this may be related to
certain features of the low temperature magnetization curves of the
spinel chromites, \znaf, \cdaf,\, and \hgaf.  A very simple Einstein
phonon model predicts a half-magnetization plateau with a unique ground
state spin configuration.  Given the known presence of large
spin-lattice coupling in these materials,\cite{Ueda:prb06,Ueda:prl05}
this seems the most likely explanation of the plateau state.  Neutron
scattering measurement of the R-state structure on the plateau in any of
these materials would provide good support of this proposal.

Nevertheless, the same ground state {\sl could} in principle be
stabilized by other sorts of interactions.  For instance, as we have
shown, further neighbor exchange with $J_3>J_2$ would also lead to the
R-state.  This relation amongst exchanges seems contrary to simple
expectations that antiferromagnetic superexchange decays with
distance.  However, the exchange paths for these two interactions (and
indeed some further-neighbor exchanges) are quite similar (see
Ref.\onlinecite{Baltzer:pr66}), so one should have an open mind to
this possibility.  It would thus be desirable to have an independent
comparison of the two theoretical models.  

One further check on the Einstein model would be to consider its
predictions in zero field.  As we have seen, despite the selection of
a unique ground state on the plateau, in zero field the Einstein model
continues to predict a large degeneracy of states.  In reality, of
course, this degeneracy will be broken by further interactions (e.g
more complex phonons, further-neighbor exchange, Dzyaloshinskii-Moriya
interactions) beyond the Einstein model.  Interestingly, the zero
field ordered phases for \znaf,\,\cdaf, and \hgaf\, are known from
neutron scattering and are known to be {\sl
  different}.\cite{chung:prl05,Lee:nat02,Lee:prl00} This is indeed
consistent with the Einstein model in the sense that the further very
small interactions beyond the model would be expected to select
different states in each material.
Interestingly, the Einstein model suggests that despite this panoply
of phases at zero field, all these materials may display a {\sl
  universal} ordering on their magnetization plateau: the R-state
structure. \cite{Bergman:prl05}

As we have seen in Sec.\ref{sec:J23}, under some circumstances
($J_3>J_2$ and $b$ not too large), the further-neighbor exchange model
may have a non-collinear ground state. This would be clearly
distinguishable from the collinear states preferred by the
spin-lattice interactions. For instance, Goldstone's theorem implies
that a collinear ground state in zero field will have two gapless spin
wave modes, while a non-collinear state will have {\sl three}.
However, we caution that additional effects beyond the Einstein model,
especially Dzyaloshinskii-Moriya interactions, could induce some small
non-collinearity even if the predominant interactions are of the
spin-lattice type.

Another general prediction of the Einstein phonon model is that the
interactions ($b$) which stabilize the plateau are {\sl of the same
  order} as those ($b'$) which select the ordered R-state out of all
possible plateau configurations.  This has physical consequences.
Specifically, the energy cost of a spin excitation which leads to a
deviation from the quantized plateau magnetization is expected to be
of the same order as a spinless excitation which re-arranges the 3:1
configurations but leaves the magnetization unchanged.  On heating the
sample, the former excitations are responsible for the rounding of the
plateau, while the latter are responsible for the destruction of the
R-state magnetic order, i.e the restoration of the spinel space group
symmetry.  Because both excitations will be excited roughly equally,
we expect that the thermal phase transition from the R-state to the
high-temperature phase should occur at a critical temperature $T_c$
which is of the same order of magnitude as the scale at which the
plateau forms, and the 3:1 constraint itself is rather strongly
violated.

Because the 3:1 constraint is not significant at this temperature, a
conventional Landau-Ginzburg-Wilson analysis of this critical point is
valid (see Ref.[\onlinecite{Bergman:prb05}] for a discussion of the
alternative scenario which would apply if $T_c \ll T_p$, where $T_p$ is
the temperature at which the plateau forms).  The result of such an
analysis -- detailed in Appendix~\ref{app:LG_theory} -- is that the
thermal transition should be {\sl first order}.  This is in agreement
with experimental findings.

There are many open directions for future work.  An important one is
to understand more microscopically the mechanisms of exchange
interactions and spin-lattice coupling.  Goodenough-Kanamori analysis
actually predicts a competition between two processes effecting the
nearest neighbor exchange interaction: antiferromagnetic {\sl direct}
Cr-Cr exchange, and ferromagnetic Cr-X-Cr superexchange in the ideal
crystal structure, because of 90$^\circ$ Cr-X-Cr (X=O,S,Se)
bonds.\cite{Baltzer:pr66} The angle of the Cr-X-Cr bonds is generally
not exactly 90$^\circ$ however, and is affected by the $u$-parameter
in the spinel structure.  The prevailing belief is that the
competition of these two processes is most strongly effected by the
overall expansion/contraction of the lattice, but the different
$u$-parameters in different materials may also be important.  Longer
distance super-exchange processes between second, third, and fourth
neighbor pyrochlore sites involve comparable exchange paths, and their
relative magnitudes are not presently clear.  Turning to spin-lattice
coupling, an interesting speculation is that the important phonon
modes are those which modify the Cr-X-Cr angles, thus strongly
affecting the nearest-neighbor superexchange contribution.  This is an
appealing possibility given the observation of changes between
ferromagnetic and antiferromagnetic behavior in small changes of
temperature and field in HgCr$_2$S$_4$.\cite{Tsurkan:06}  A broader understanding of the microscopic
mechanisms of spin and spin-lattice interactions in these materials
would be of considerable interest even beyond the realm of frustrated
magnetism. 

In summary, the study here clearly highlights the sensitivity of the
magnetic state in antiferromagnetic pyrochlores to further microscopic
interactions.  A simple Einstein model makes the prediction that the
R-state order should be observed in many materials and we hope that
detailed neutron scattering studies will be forthcoming to test this
prediction.

\acknowledgments

We are grateful to H. Ueda, Y. Motome, M. Matsuda, H. Takagi, and Y.
Ueda for discussions, and to H. Ueda, H. Mitamura, T. Goto, and H.
Takagi for sharing their experimental data with us prior to publication.
We also wish to acknowledge fruitfull discussions with S. Trebst regarding 
some of the numerical work presented in this paper.
This work was supported by NSF Grant DMR04-57440, PHY99-07949, and the
Packard Foundation.

\appendix

\section{Landau-Ginzburg Theory for the finite temperature transition out of the plateau state}
\label{app:LG_theory}

In the R-state\cite{Bergman:prl05} the enlarged spin periodicity, including 4 unit cells of the underlying pyrochlore lattice,
manifests itself in the appearance of non-zero Fourier components with a momentum vector not at the $\Gamma$ point in the Brillouin Zone (BZ). These Fourier components therefore serve as order parameters for the R-state.  
Then, using these order parameters
and their transformation under the various lattice symmetries, 
we can construct a Landau-Ginzburg theory to determine whether a finite temperature transition 
out of this ordered state should be 1st order or 2nd order, at least at the mean field level.

Previous work by some of the authors\cite{Bergman:prb05} has encountered the same R-state in a different formulation, using a PSG (Projective Symmetry Group) analysis. We find the Fourier components are at the 3 momentum vectors 
${\bf k}_1 = \frac{1}{a} \left( \pi, 0, 0 \right)$,
${\bf k}_2 = \frac{1}{a} \left( 0, \pi, 0 \right)$,
and 
${\bf k}_3 = \frac{1}{a} \left( 0, 0, \pi \right)$.
These form a k-star of the point symmetry group of the lattice. This particular k-star has only 6-dimensional irreps (irreducible representations) of the point symmetry group. Therefore the order parameter can be cast as a 6-component real vector. These components can be understood as degrees of freedom equivalent to the real and imaginary parts of the 3 Fourier components.

The R-state corresponds to 8 configurations of this more general order parameter
$
{\vec v} = \left( \sigma_1,\sigma_2,\sigma_3,0,0,0 \right) v
$
or
$
{\vec v} = \left( 0,0,0,\sigma_4,\sigma_5, \sigma_6 \right) v
$
where $\sigma_j = \pm 1$, and we allow only the $8$ cases where
$\sigma_1 \sigma_2 \sigma_3 = +1$ or
$\sigma_4 \sigma_5 \sigma_6 = +1$.

This order parameter was derived using the PSG analysis. However, it is possible to derive this order parameter more directly, by finding the ``density'' of minority sites $\rho_j =\frac{1}{2} \left(1 - S^z_j\right) $.  A straightforward Fourier analysis results in
\begin{equation}
\begin{split} &
\rho(x,y,z) =  
\\ & \frac{1}{4} 
\Big( 1 + \sigma_4 \sqrt{2} \cos{\frac{\pi}{a}x} + \sigma_5 \sqrt{2} \cos{\frac{\pi}{a}y} + \sigma_6 \sqrt{2} \cos{\frac{\pi}{a}z}
\\ &
- \sigma_1 \sqrt{2} \sin{\frac{\pi}{a}x} - \sigma_2 \sqrt{2} \sin{\frac{\pi}{a}y} - \sigma_3 \sqrt{2} \sin{\frac{\pi}{a}z}
\Big)
\; ,
\end{split}
\end{equation}
where $(x,y,z)$ are the real space coordinates of the pyrochlore lattice sites.

In this representation we construct invariants (or Casimir operators) from the 6 components $\{v_j\}_{j=1}^6$ of 
${\vec v}$. The Landau-Ginzburg (LG) theory is then constructed out of all these invariants.
In this way we find the most general LG free energy allowed by the symmetry of the lattice
\begin{equation}
\begin{split}
F = & m {\vec v}^2 + \gamma \left( v_1 v_2 v_3 + v_4 v_5 v_6 \right) + u_1 {\vec v}^4
\\ & 
+ u_2 \sum_{j=1}^3 v_j^2  \sum_{i=4}^6 v_i^2 
\\ &
+ u_3 \left[ \sum_{i \neq j =1}^3 \left( v_i v_j \right)^2 + \textrm{other trio} \right]
\; .
\end{split}
\end{equation}
The 8 R-state configurations are favored for a part of the coupling space
$\gamma<0, m<0, u_1>0, u_2>0$, and $u_3<0$.   

The transition into the orderless state is tuned by $m$ changing sign. We take 
$
{\vec v} = \left( 1,1,1,0,0,0 \right) v
$
without loss of generality, and simplify the free energy to a single variable function
\begin{equation}
F(v) = 
m' v^2 - |\gamma| v^3 + |u'| v^4
\; .
\end{equation}
Due to the cubic term, the transition is predicted by MFT to be 1st order.

\section{Proof of bending rule inducing the R-state }
\label{app:bending}

In this appendix we explain how from the ``bending'' rule for the $3:1$ configurations, 
we can construct only the R state.

Consider the unit cell of the R state, including 4 up pointing tetrahedra as in Fig.~\ref{fig:Pyro_cell1}.
We will show that using bending configurations on pairs of tetrahedra, we will find a unique state 
(up to lattice symmetries).

\begin{figure}[hbt]
	\centering
	\subfigure[]{
	  \label{fig:proof_1}
    \includegraphics[height=1.5in]{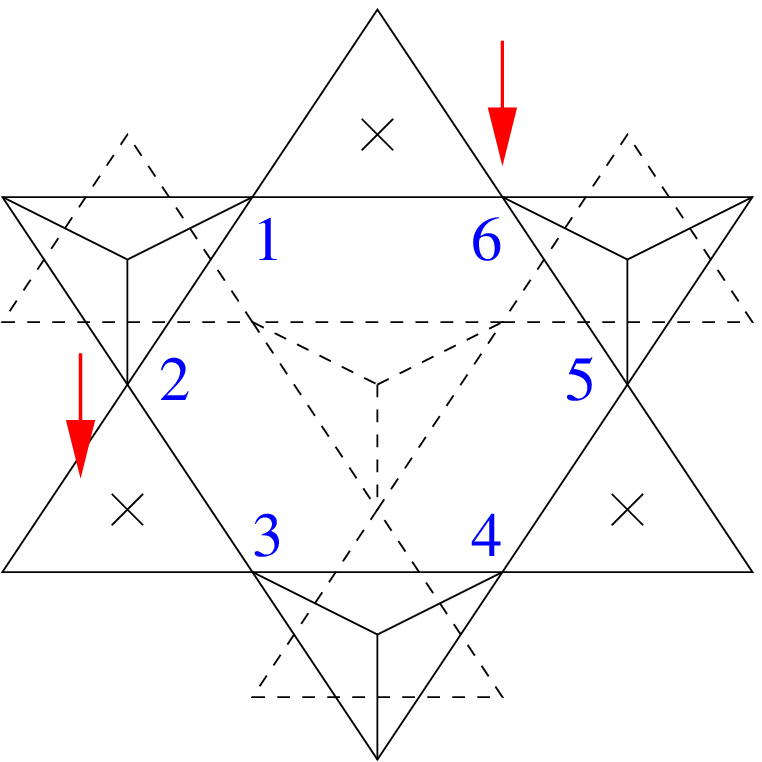}}
	\subfigure[]{
	  \label{fig:proof_2}
    \includegraphics[height=1.5in]{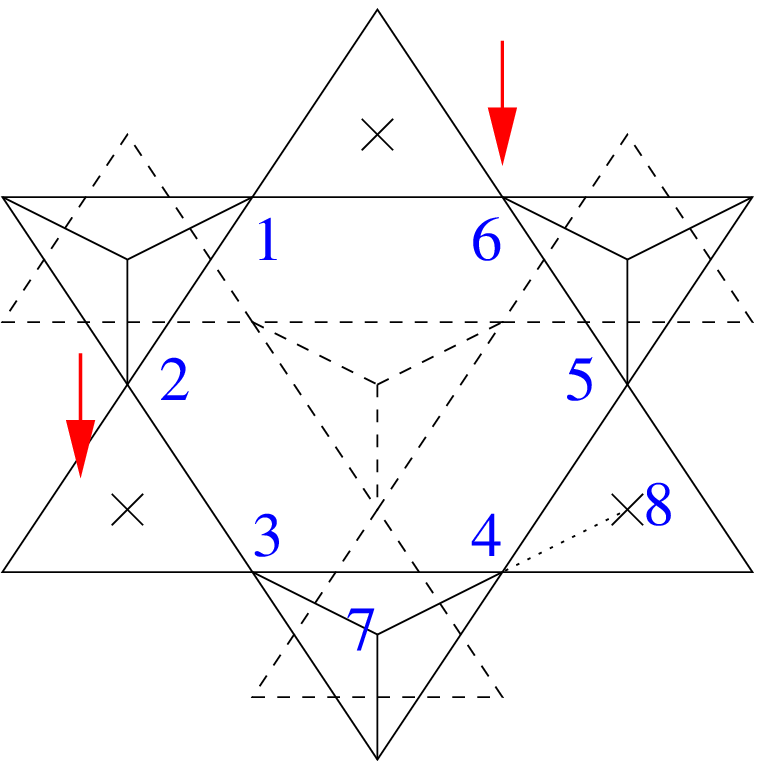}}
	\subfigure[]{
	  \label{fig:proof_3}
    \includegraphics[height=1.5in]{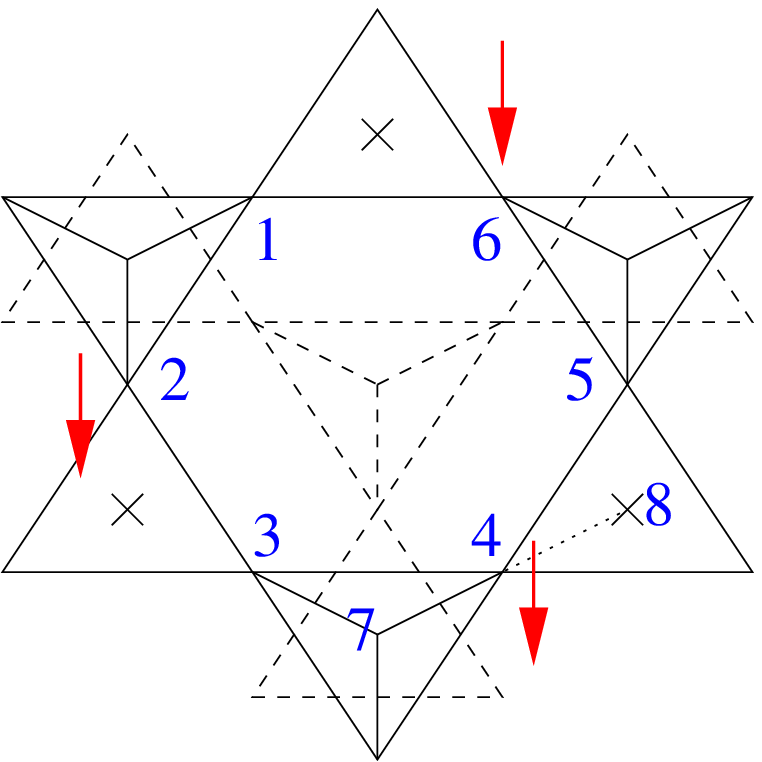}}
	\subfigure[]{
	  \label{fig:proof_4}
	  \includegraphics[height=1.5in]{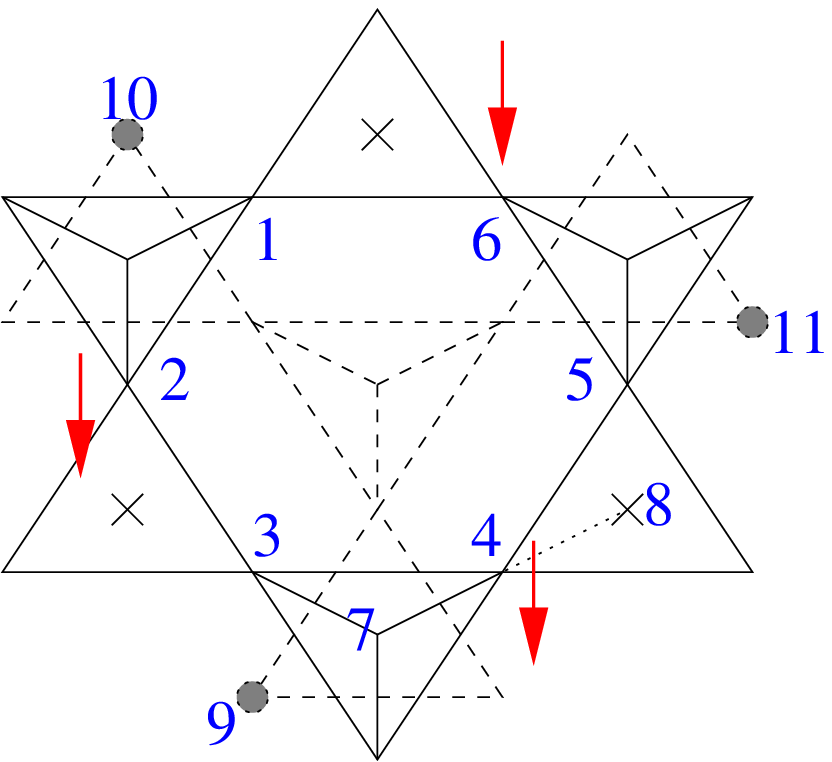}}
	\subfigure[]{
	  \label{fig:proof_5}
	  \includegraphics[height=1.5in]{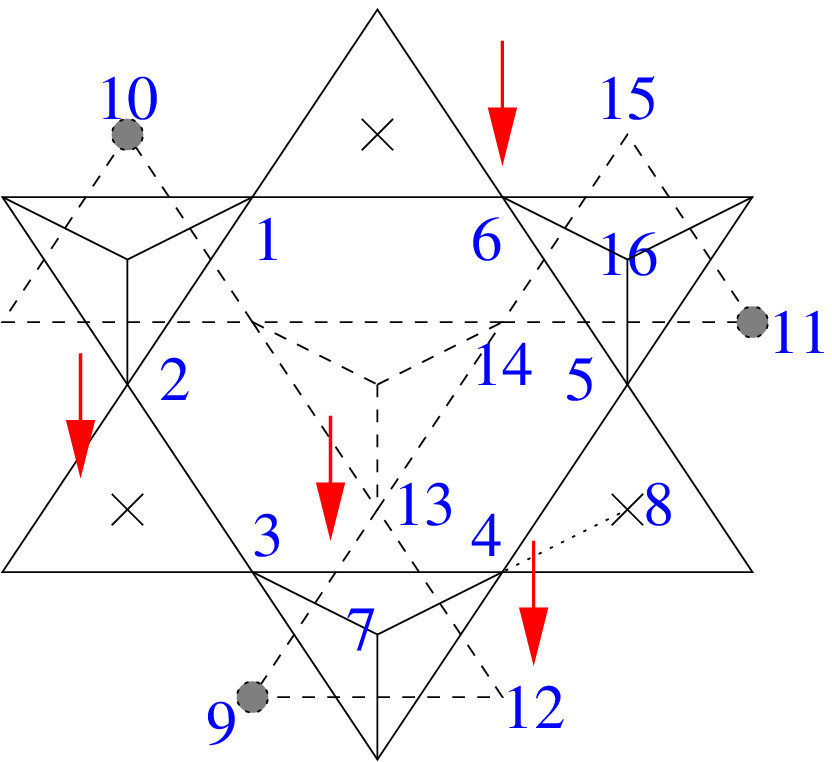}}
  \subfigure[]{
	  \label{fig:proof_6}
	  \includegraphics[height=1.5in]{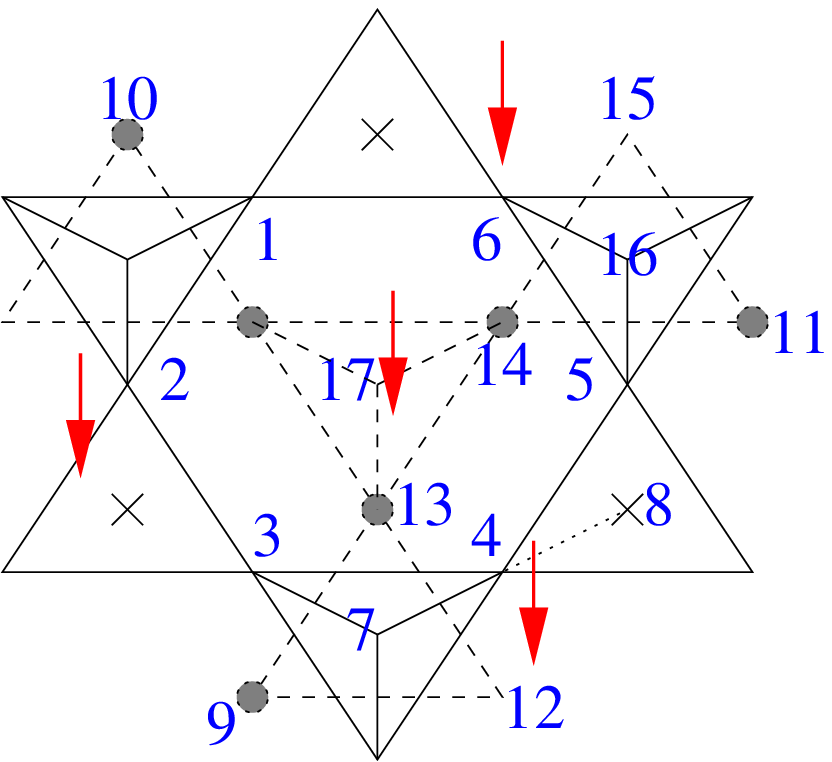}} 
	\caption{(Color online) Stages in the graphical proof that the bending rule allows only the R state to form with the $3:1$ constraint. The down pointing triangles, with lines joining at their centers, represent tetrahedra pointing out of the plane of the paper, with the center point representing the upper corner. The up pointing triangles in the figures represent tetrahedra pointing into the paper, with the lowest corner being represented by the center position (sometimes marked by a small cross). The dashed lines represent an upper layer of tetrahedra, above the solid line tetrahedra. It is instructive to contrast this projection with Fig.~\ref{fig:Pyro_cell1} - both show the \emph{same} cluster of tetrahedra. To avoid clutter, we mark only the minority sites, by down pointing (red) arrows. We mark sites that we conclude are not allowed to be minority sites by a full gray circle.}
	\label{fig:proof}
\end{figure}

We start by picking a pair of tetrahedra to be in the bending configuration. We can then pick the unit 
cell orientation such that it matches the placements of the minority sites in Fig.~\ref{fig:proof_1} - the two
adjacent tetrahedra share site $1$ and the two minority sites are $2$ and $6$.

Next we consider the tetrahedron pairs sharing sites $3$ and $5$, shown in Fig.~\ref{fig:proof_2}. 
For the pair sharing site $3$, we can pick sites $4$ or $7$ to be minority sites. Similarly, 
for the pair sharing site $5$, we can pick sites $4$ or $8$ to be minority sites. We cannot 
pick $4$ and $7$ or $4$ and $8$, since then we will have nearest neighbor
minority sites. We also cannot pick $7$ and $8$, since then the pair of tetrahedra sharing site $4$ will \emph{not}
be in a bending configuration - sites $7$,$4$ and $8$ sit on a straight line. We must therefore choose site $4$ to be 
a minority site, (see Fig.~\ref{fig:proof_3}). Already a tendency to form flippable plaquettes is evident.

Now we turn to the upper layer of tetrahedra (the tetrahedra outlined by dashed lines). Considering the pair
of tetrahedra adjoined at site $7$, we conclude that site $9$ \emph{cannot} be a minority site. Similar considerations
deem sites $10$ and $11$ cannot be minority sites. This should be most clearly evident from the 3-fold rotational 
symmetry of the minority site configuration about an axis perpendicular to the paper. We mark these sites 
in Fig.~\ref{fig:proof_4} by a full gray circle.

The pair of tetrahedra sharing site $7$ can allow a minority site on either site $12$ or $13$. If we choose 
site $13$ (as in Fig.~\ref{fig:proof_5}), then now we cannot choose site $14$ to be a minority site, 
as it neighbors site $13$, and we also 
cannot choose site $15$ to be a minority site, as $13$,$14$ and $15$ sit on a straight line. Now considering the pair
of tetrahedra sharing site $16$, we reach an impasse - the tetrahedron of sites $11$, $14$, $15$ and $16$
cannot have a minority site on any one of its corners!  Therefore, we \emph{cannot} choose choose $13$
to be a minority site, we can only choose site $12$!
 
Due to the 3-fold rotation symmetry of the spin structure we have layed out so far 
the same argument applies to all three down pointing tetrahedra in the upper (dashed) layer.
The single up pointing tetrahedron must therefore have a minority site located at site $17$.
The resulting configuration, shown in Fig.~\ref{fig:proof_6} is the unit cell of the R-state.

\begin{figure}[hbt]
	\centering
	\subfigure[]{
	  \label{fig:proof_7}
    \includegraphics[height=2.0in]{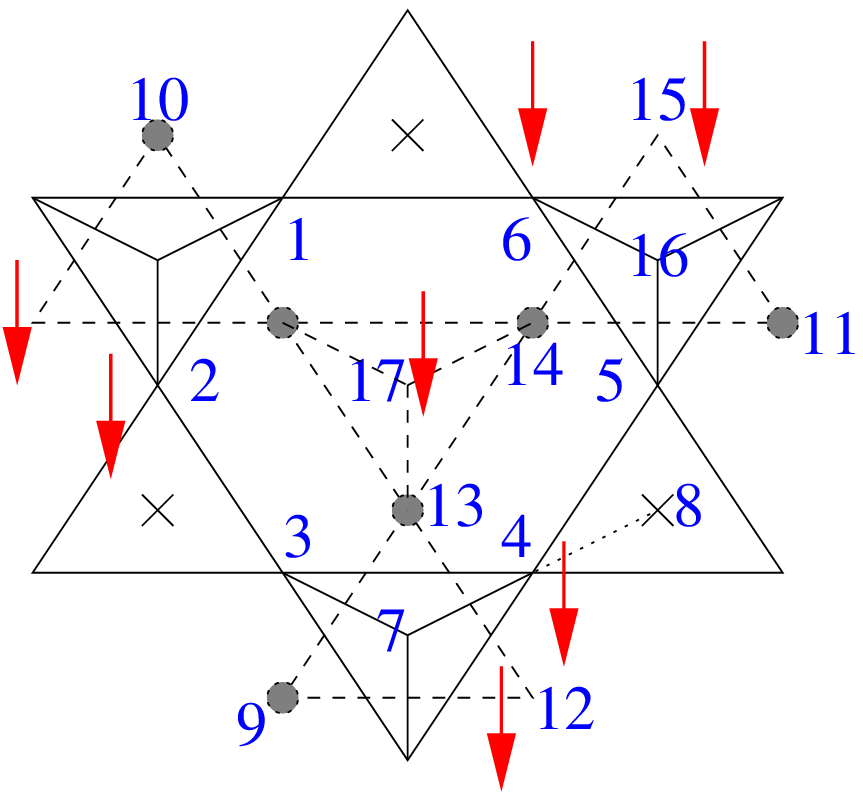}}
	\subfigure[]{
	  \label{fig:proof_8}
    \includegraphics[height=2.0in]{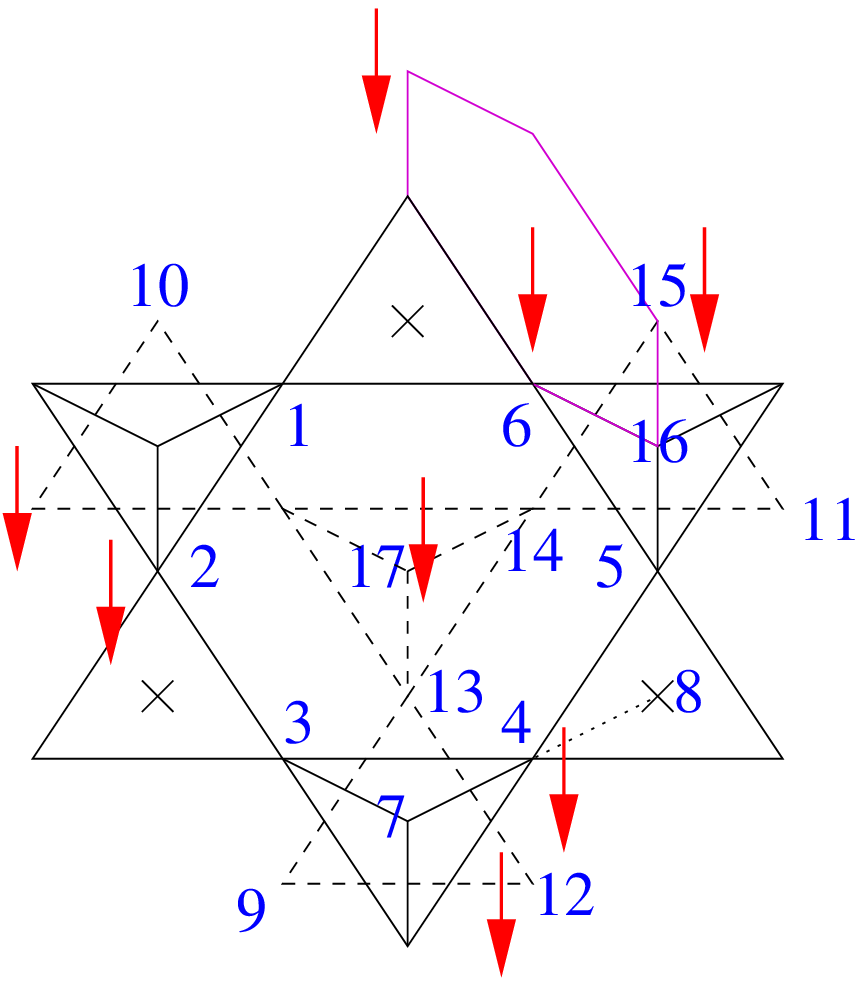}}
	\caption{(Color online) Stages in the graphical proof that the bending rule allows only the R state to form with the $3:1$ constraint. The down pointing triangles, with lines joining at their centers, represent tetrahedra pointing out of the plane of the paper, with the center point representing the upper corner. The up pointing triangles in the figures represent tetrahedra pointing into the paper, with the lowest corner being represented by the center position (sometimes marked by a small cross). The dashed lines represent an upper layer of tetrahedra, above the solid line tetrahedra. It is instructive to contrast this projection with Fig.~\ref{fig:Pyro_cell1} - both show the \emph{same} cluster of tetrahedra. To avoid clutter, we mark only the minority sites, by down pointing (red) arrows. We mark sites that we conclude are not allowed to be minority sites by a full gray circle.}
	\label{fig:cells}
\end{figure}

Next we demonstrate that the R state is the only $3:1$ spin configuration that can be constructed from these cell structures.
We can add the minority site positions on the remaining tetrahedra in the cell cluster, as shown in Fig.~\ref{fig:proof_7}.
Each cell has $4$ hexagonal plaquettes. One is ``flippable'' - alternating between majority and minority sites. The other 
three plaquettes have one minority site, and $5$ majority sites. There is only one way to attach an identical cell with the flippable plaquette, but apriori there are three ways to attach two cells through a plaquette with only $1$ minority site
(we have the freedom to choose which one of the other $3$ plaquettes of the second cell is the flippable one).

Consider the cell sharing the hexagonal plaquette that involves sites $1,6,16$ and $14$. Since sites $6$ and $15$ are already set to be minority sites, the flippable plaquette in this cell \emph{must} be the one going through sites $6,16$ and $15$,
marked by purple lines. We therefore do \emph{not} have three choices of how to place the minority sites on this cell, 
but rather only one. The whole cluster, with the given spin configuration, still retains a 3-fold rotational 
symmetry about an axis perpendicular to the paper. We use this symmetry to deduce that the same considerations 
apply to the other three plaquettes with one minority site, and there is no freedom in how to choose the spin 
configurations in \emph{all} the surrounding cells. Therefore, there can only be one way to arrange the minority 
sites with the given cell, and that is the R-state configuration.

\end{document}